\documentstyle[11pt]{article}

\begin{document}
\begin{center}
{\Large Sea quark content of nucleons from proton-induced
Drell-Yan production}

\vspace{10mm}
\renewcommand{\thefootnote}{\fnsymbol{footnote}}
{\normalsize Bo-Qiang Ma\footnote{
Fellow of Alexander von Humboldt Foundation,
on leave from
Institute of High Energy Physics, Academia Sinica, P.O.Box 918(4),
Beijing 100039, China},
Andreas Sch\"afer, and Walter Greiner}

\vspace{8mm}
{\large
        Institut f\"ur Theoretische Physik der
        Universit\"at Frankfurt am Main, Postfach 11 19 32,
        D-60054 Frankfurt, Germany}
\vspace{4mm}

\end{center}

{\large \bf Abstract }
We analyse the proton-induced Drell-Yan production in both the flavor
asymmetry and isospin breaking explanations for the violation of the
Gottfried sum rule. Consequences from three different forms of corrections
to the flavor and isospin symmetric parton distributions are examined.
It is found that the calculated results are sensitive to the choices
of parameters and parton distributions, and to the ways the corrections
are introduced. All three forms of corrections could be consistent with
the recent Fermilab Experiment E772 data for the ratio of cross section
$R=\sigma_{W}/\sigma_{IS}$ and for the shape of the differential cross
section $m^{3}d^{2}\sigma/dx_{F}dm$ for $^{2}$H.

\vspace{2mm}

\break

Recently, the violation of the Gottfried sum rule (GSR) reported
by the New Muon Collaboration (NMC) \cite{NMC91}
has received attention
and a number of papers have been devoted to discuss the
explicit form of flavor sea distributions in the nucleons\cite{Ma93}.
Several different explanations for the
origin of the GSR violation have been proposed,
such as a flavor asymmetry of the nucleon sea\cite{Pre91},
isospin symmetry breaking between the proton
and the neutron\cite{Ma92},
non-Regge behaviors at small $x$ \cite{Mar90},
and nuclear
effects like the mesonic exchanges in the deuteron \cite{Kap91}
and nuclear bindings \cite{Epe92}.
The possibilities to discriminate
between these different explanations through other
processes have also been analysed
\cite{Ma93},\cite{Ell91}-\cite{Epe92b}.
It was first pointed out
by Ellis and Stirling \cite{Ell91} that
proton-induced
Drell-Yan production is one of the sensitive processes
that can provide some further information concerning the
origin of the GSR violation. The Drell-Yan asymmetry
$A_{DY}(x_{1},x_{2})=
\frac{\sigma_{pp}(x_{1},x_{2})-\sigma_{pn}(x_{1},x_{2})}
{\sigma_{pp}(x_{1},x_{2})+\sigma_{pn}(x_{1},x_{2})}$
for $x_{1}=x_{2}$
was introduced and it was found that this quantity
can change sign depending on whether the sea is flavor symmetric
or not. The shape of the differential
cross section was observed to be different for pp collisions
and pn collisions. Kumano and Londergan examined the
Drell-Yan asymmetry for several different quark distributions
assuming flavor asymmetry
and showed that the estimates for this quantity differ
widely\cite{Kum92}.
The modification to the value of
$A_{DY}$ from nuclear binding effects was found to
be very small in comparison with the results in
the flavor and isospin symmetry case\cite{Epe92b}.
It has also been indicated by us \cite{Ma93}
that $A_{DY}$ has approximately
the same value for the flavor asymmetry and isospin breaking
explanations,
thus it is difficulty to distinguish between
these two explanations.

The recent proton-induced Drell-Yan production data reported by
Fermilab Experiment E772 \cite{McG92}  show
no evidence for
a large flavor asymmetric sea.  However, in a most recent
study it was shown that a sea flavor asymmetric parton
distribution derived from a chiral quark model can provide
a satisfactory description of the E772 data\cite{Eic93}.
Thus we need to study further the implications of the
E772 data for the explicit flavor distributions in the quark sea
of nucleons.
We will examine in this paper three different forms
of corrections to the flavor and isospin symmetric parton distributions
in the flavor asymmetry and isospin breaking
explanations.
We will show that the results for a hard correction
$a(1-x)^{b}$ (i.e., a non-Pomeron form, referred to as I)
and for a medium-hard correction $ax^{-1/2}(1-x)^{b}$
(i.e., a pionic form, referred to as II)
are rather sensitive
to the parameters used, to the choices of parton distributions, and
to the ways
the corrections are introduced. They could be consistent
with the E772 data for the ratio of cross section
per nucleon $R=\sigma_{W}/\sigma_{IS}$
and
for the shape
of the differential cross section $m^{3}d^{2}\sigma/dx_{F}dm$ for
$^{2}$H.
The results
for a soft
correction $ax^{-1}(1-x)^{b}$  (i.e., a Pomeron form, referred to as
III) are less
sensitive to parameters and the ways the corrections are
introduced, and
can describe the E772 data.
Thus the available Drell-Yan data pose constraints
on the explicit form of the sea parton distributions and on
how the corrections to
the flavor and isospin symmetric sea distributions
are introduced in the two explanations,
but cannot rule out any possible explanations.

We show that the Ellis-Stirling sign criterion \cite{Ell91}
of the Drell-Yan Asymmetry
$A_{DY}(x_{1},x_{2})$
is also correct in the case $x_{1}\neq x_{2}$.
The expression for the cross section is
\begin{equation}
\sigma^{pN}(x_{1},x_{2})=
\frac{d^{2}\sigma^{pN}}{d x_{1}d x_{2}}=
\frac{4\pi\alpha^{2}}{9m^{2}}K
\sum_{i}e_{i}^{2}[q_{i}^{p}(x_{1})
\overline{q}_{i}^{N}(x_{2})+
\overline{q}_{i}^{p}(x_{1})
q_{i}^{N}(x_{2})],
\end{equation}
where $x_{1}$ and $x_{2}$ are the Bjorken variables for the beam
hadron (p) and the target hadron (N), $m^{2}=Sx_{1}x_{2}$ with
$m$ denoting the mass of the dilepton and $\sqrt{S}
=\sqrt{2M_{p}(E_{lab}+M_{p})}$ denoting the
total energy of the beam and the target in the center of mass frame,
and $K$ is a normalization factor due to complicated higher-order
processes.
We thus obtain, in the flavor asymmetry explanation,
\begin{equation}
A_{DY}(x_{1},x_{2})=
\frac{
(4\overline{u}_{1}-\overline{d}_{1})(u_{2}-d_{2})
-(\overline{d}_{2}-\overline{u}_{2})(4u_{1}-d_{1})}
{(4\overline{u}_{1}+\overline{d}_{1})(u_{2}+d_{2})
+(\overline{d}_{2}+\overline{u}_{2})(4u_{1}+d_{1})
+4\overline{s}_{1}\overline{s}_{2}},
\label{eq:adyud}
\end{equation}
and, in the isospin breaking explanation,
\begin{equation}
A_{DY}(x_{1},x_{2})=
\frac{
3\overline{q}^{p}_{1}(u_{2}-d_{2})-
(\overline{q}^{n}_{2}-\overline{q}^{p}_{2})
(4u_{1}+d_{1}+5\overline{q}^{p}_{1})}
{5\overline{q}^{p}_{1}(u_{2}+d_{2}+
\overline{q}^{n}_{2}-\overline{q}^{p}_{2})+
(\overline{q}^{p}_{2}+\overline{q}^{n}_{2})
(4u_{1}+d_{1})+4\overline{s}_{1}\overline{s}_{2}},
\label{eq:adypn}
\end{equation}
where $u_{1}=u(x_{1})$, $u_{2}=u(x_{2})$, $d_{1}=d(x_{1})$
and $d_{2}=d(x_{2})$ {\it et al.} are the quark distributions
in the proton.
It can be easily found that the quantity $A_{DY}(x_{1},x_{2})$
is always positive
in the flavor and isospin symmetry case (as $u_{2}/d_{2}\geq1$)
whereas it can change sign for the flavor and/or isospin
asymmetry cases.
The ratio $R=\sigma_{W}/\sigma_{IS}$ can be expressed as
\begin{equation}
R=\frac{N\sigma^{pn}+Z\sigma^{pp}}{A/2\;(\sigma^{pn}+\sigma^{pp})}
=1-\frac{N-Z}{A}A_{DY}.
\label{eq:ady}
\end{equation}
This quantity is always smaller than unity for the flavor and
isospin symmetry case whereas
it is possible to get values above unity for the flavor
and/or isospin asymmetry cases.
Therefore a confirmation of any point larger than unity will
be the evidence for flavor asymmetry in the nucleon sea or
the isospin breaking between the proton and the neutron.
The consideration of further nuclear effects
could complicate the above
analysis.
To calculate the differential cross section $m^{3}d^{2}\sigma/dx_{F}dm$
for $^{2}$H,
we use the expression
\begin{equation}
m^{3}\frac{d^{2}\sigma^{AB}}{dx_{F}dm}
=\frac{8}{9}\pi\alpha^{2}
(\frac{x_{1}x_{2}}{x_{1}+x_{2}})K\sum_{i}e_{i}^{2}
[q_{i}^{A}(x_{1})\overline{q}_{i}^{B}(x_{2})
+\overline{q}_{i}^{A}(x_{1})q_{i}^{B}(x_{2})],
\end{equation}
where
$x_{F}=x_{1}-x_{2}$
and the $K$ factor is adjusted
to fit the large-$x_{F}$ data.

We first modify the sea by
\begin{equation}
\begin{array}{clcr}

\overline{d}(x)=\overline{q}_{0}(x)+5/6 \:\Delta (x);\\

\overline{u}(x)=\overline{q}_{0}(x)-5/6 \:\Delta (x),
\label{eq:mdfa1}
\end{array}
\end{equation}
as adopted in Ref.~\cite{Ell91},
in the flavor asymmetry explanation.
In the isospin breaking case we modify the sea by
\begin{equation}
\begin{array}{clcr}

\overline{q}^{n}(x)=\overline{q}_{0}(x)+ \Delta (x);\\

\overline{q}^{p}(x)=\overline{q}_{0}(x).
\label{eq:mdib}
\end{array}
\end{equation}
We adopt the three forms of corrections as mentioned above for
$\Delta(x)=\overline{q}^{n}(x)-\overline{q}^{p}(x)
=\frac{3}{5}(\overline{d}(x)-\overline{u}(x))$
in the two explanations respectively,
with the parameters $a$ and $b$ adjusted
to fit $\int_{0}^{1} dx \Delta(x)=0.084$ as required to
reproduce the observed NMC result
$S_{G}=0.240$.
We indicate that all three forms of corrections
could be compatible with the NMC data $F_{2}^{n}(x)/F_{2}^{p}(x)$,
from which the Gottfried sum $S_{G}=0.240$ is obtained.
The reason is that there are many different parametrizations
of the parton distributions which may give different
detailed features.
In Ref.~\cite{Pre91} a
correction of form I
was suggested to represent the difference between the
up and down quark sea in the flavor asymmetry explanation,
based on an old parametrization MRS(B)\cite{Mar88}.
The corrections of form II have been studied by many
authors in a framework where the flavor asymmetry in the sea
is attribute to
the excess of $p \rightarrow n+\pi^{+}$ over
$p \rightarrow \Delta^{++}+\pi^{-}$ (or $\pi^{+}$ over
$\pi^{-}$)\cite{Pi},
or from a more microscopic point of view to the excess of
$u\rightarrow d+\pi^{+}$ over $u \rightarrow u+\pi^{0}$
and $d\rightarrow u+\pi^{-}$ over
$d\rightarrow d+\pi^{0}$\cite{mPi}.
Considering that the sea parton distributions are of Pomeron form,
we can attribute form III correction as a small perturbation
in the sea distributions.
In Fig.~1 we present the results of the calculated
ratio $F_{2}^{n}/F_{2}^{p}$ by using three forms of corrections
based on two parametrizations of parton distributions, i.e.,
EHLQ set 1 \cite{Eic84} and
the new MRS(S0) set\cite{Mar93}.
We see that form I and II corrections
give good descriptions of the NMC data based on the old
parametrization EHLQ set 1.
The correction of form III is also consistent with
the data based on the new parametrization
MRS(S0) set. Changing the parameter $b$ causes very small
changes in $F_{2}^{n}/F_{2}^{p}$.
Therefore we can examine the influences from the
three form corrections in the description of
the ratio $R=\sigma_{W}/\sigma_{IS}$ and the shape of
$m^{3}d^{2}\sigma^{pd}/dx_{F}dm$ in Drell-Yan process,
regardless of the detailed parametrization dependence.

In Figs.~2, 3 and 4 we compare the
calculated results of
$R=\sigma_{W}/\sigma_{IS}$
and $m^{3}d^{2}\sigma^{pd}/dx_{F}dm$
with the E772 data.
The data of $R$ at large $x_{target}$ do not exclude
points larger than unity. The data at small $x_{target}$
have been corrected to remove the nuclear shadowing
effect\cite{McG92}.
Because W is significantly heavier than C and Ca, the targets used to
determine the shadowing factor $\alpha_{sh}$, we can not exclude the
possibility that the shadowing correction is larger than
estimated by using $\sigma_{A}=\sigma_{N}A^{\alpha_{sh}}$. This could
increase the data slightly at small $x_{target}$.
{}From Figs.~2 and 3 we find large parameter
dependence in the calculated results
for corrections of form I and II.
For small $b$ we see that the calculated
$R$ is larger than the E772 data in the flavor asymmetry explanation
whereas it is compatible with the data in the isospin
breaking explanation.
$R$ is slightly dependent on $b$ in the isospin breaking case
whereas it
decreases significantly for large $b$
in the flavor asymmetry case.
The value of $m^{3}d^{2}\sigma^{pd}/dx_{F}dm$  will decrease at
negative $x_{F}$ for large $b$ in the two explanations.
Because the shape of the
calculated $m^{3}d^{2}\sigma^{pd}/dx_{F}dm$ is dependent on the
the parton distributions, we can hardly say that this
trend is in disagreement with the data.
{}From Figs.~2 and 3 we see that the calculated results
of $R$ and $m^{3}d^{2}\sigma^{pd}/dx_{F}dm$
could be
consistent with both the data depending on the parametrization we
choose. For example, we can list a number of results which are
consistent with the data, e.g., form I correction
with $b=12.2$ using the MRS(S0) set
parton distributions, form II correction
with $b=6.68$ using the MRS(S0) set parton
distributions, form I correction in the isospin
breaking explanation with $b=5.02$ using the EHLQ set 1 parton
distributions, etc..
However, the calculated results are not so sensitive
for form III corrections.
{}From Fig.~4
we see that the calculated $R$ and $m^{3}d^{2}\sigma^{pd}/dx_{F}dm$
are not so different from those in the flavor and isospin
symmetry case, thus they are consistent with the data for both sets of
parton distributions.

{}From the above discussions we see that for form I and II
corrections there are large
differences between the results in the two explanations.
This feature, which disagrees with our previous conclusion \cite{Ma93}
that the Drell-Yan asymmetry
$A_{DY}$ has approximately
the same value for the two explanations, is due to the fact that
the assumption
$4\overline{u}-\overline{d} \sim (\overline{q}^{p}+8\overline{q})/3$
made in Ref.~\cite{Ma93}
is not satisfied for the corrections Eqs.~(\ref{eq:mdfa1})
and (\ref{eq:mdib}). If we modify the sea by
\begin{equation}
\begin{array}{clcr}

\overline{d}(x)=\overline{q}_{0}(x)+5/3 \:\Delta (x);\\

\overline{u}(x)=\overline{q}_{0}(x)
\label{eq:mdfa2}
\end{array}
\end{equation}
in the flavor asymmetry explanation, we can reduce the difference
between the results in the two explanations significantly. This can be
seen from Fig.~5, where the results for form I
correction with $b=5.02$ are presented. We see that the
results for $R$ in the two explanations
are approximately the same and are consistent with the data.
The calculated $m^{3}d^{2}\sigma^{pd}/dx_{F}dm$
increases at negative $x_{F}$ compared with that in the flavor and
isospin symmetry case, thus the shape is in agreement with
the data for the EHLQ set 1 parton distributions.
Therefore the Drell-Yan process is hardly able to settle whether
the violation of the Gottfried sum rule is due to flavor
asymmetry or isospin breaking, as
we have concluded in Ref.~\cite{Ma93}.

In summery, we analysed the proton-induced Drell-Yan
production in both the flavor asymmetry
and isospin breaking explanations
for the violation of the Gottfried sum rule.
We examined three different forms of corrections
to the flavor and isospin symmetric parton distributions,
and found that the results are sensitive to parameters,
parton distributions,
and the ways the
corrections are introduced.
We conclude that
the Drell-Yan process can
settle whether the sea is flavor and isospin symmetric or not
according to the Ellis-Stirling sign criterion, and it leads to
constraints on the explicit form of the sea parton
distributions and on
how the sea is deformed from the flavor and
isospin symmetry case.

\noindent
{\large \bf ACKNOWLEDGMENTS}

One of the authors (B.~-Q.~M) would like to acknowledge the
Alexander von Humboldt Foundation for financial support.

\break
\noindent
{\large \bf Figure Captions}
\renewcommand{\theenumi}{\ Fig.~\arabic{enumi}}
\begin{enumerate}
\item
The ratio $F_{2}^{n}/F_{2}^{p}$ as a function of the Bjorken scaling
variable $x$. The data are from the NMC measurement\cite{NMC91}.
The solid, dashed, and dotted curves are the calculations
in the flavor and isospin symmetry, flavor asymmetry, and isospin
breaking cases, respectively, with
thin and thick curves corresponding to results
based on the EHLQ set 1 \cite{Eic84}
and MRS(S0) set \cite{Mar93}
of parton distributions.
In the figure the dashed and dotted curves are in
coincidence.
(a) The results for form I
correction with $b=9.6$. (b) The results for form II correction with
$b=6.68$. (c) The results for form  III
correction with $b=6.06$.
\item
The ratio $R=\sigma_{W}/\sigma_{IS}$ as a function of the
Bjorken variable for the target $x_{target}$ and
the differential cross section $m^{3}d^{2}\sigma/dx_{F}dm$ for
$^{2}$H as a function of $x_{F}$.
The data are from Fermilab Experiment E772, Ref.~\cite{McG92}.
The curves, which have the same
meaning as those in Fig.~1, represent
the calculations for form I correction, with the
following parameter: (a) $b=9.6$; (b) $b=12.2$;
and (c) $b=5.02$. The
results of (c) in the flavor asymmetry case suffer the flaw of
having negative $\overline{u}(x_{target})$ at large $x$
for the EHLQ set 1
parton distributions.
\item
Same as Fig.~2, but the calculated results
are for form II correction, with the following parameter:
(a) $b=6.68$; (b)
$b=4.13$.
\item
Same as Fig.~2, but the calculated results are for
form III correction, with the following parameter: (a) $b=6.06$; (b)
$b=3.06$.
\item
Same as Fig.~2(c), but the results in the flavor asymmetry
case are calculated by using Eq.~(\ref{eq:mdfa2}), instead of
Eq.~(\ref{eq:mdfa1}). This removes the negative
$\overline{u}(x_{target})$ at large $x$  occurred in Fig.~2(c).
\end{enumerate}

\end{document}